# On the Evolution of Ion Bunch Profile in the Presence of Longitudinal Coherent Electron Cooling

G. Wang[1,*]

[1]*Collider-Accelerator Department, Brookhaven National Laboratory, Upton, New York 11973-5000, USA*

In the presence of longitudinal coherent electron cooling, the evolution of the line-density profile of a circulating ion bunch can be described by the 1-D Fokker-Planck equation. We show that, in the absence of diffusion, the 1-D equation can be solved analytically for certain dependence of cooling force on the synchrotron amplitude. For more general cases with arbitrary diffusion, we solved the 1-D Fokker-Planck equation numerically and the numerical solutions have been compared with results from macro-particle tracking.

## I. INTRODUCTION

The future electron-ion collider demands a strong hadron cooling technique to reach the luminosity level where all the relevant physics can be fully covered. As a potential candidate to provide such a cooling technique, the concept of Coherent electron Cooling (CeC) and its variants have been extensively investigated [1-5]. Both analytical and simulation tools have been developed to predict the ion bunch evolution in the presence of CeC, which is essential for diagnosing as well as optimizing the cooling system[6, 7]. While simulation through macro-particle tracking is the most straightforward approach in predicting the ion bunch evolution under cooling, analytical tools are needed in both benchmarking the simulation code and providing a fast estimate for the ion bunch profile.

In case that the system has only longitudinal cooling, the evolution of the ion bunch under cooling can be described by the 1-D Fokker-Planck equation. The analytical tools are being developed to solve the 1-D Fokker-Planck equation with given cooling rate and diffusion coefficient. We review the 1-D Fokker-Planck equation and its equilibrium solutions in section II. In section III, we derived an analytical solution of the 1-D Fokker-Planck equation for a specific dependence of the cooling rate on synchrotron oscillation amplitude, in the limit of vanishing diffusion coefficient. Section IV contains our approach of numerically solving the 1-D Fokker-Planck equation for finite diffusion coefficient and arbitrary dependence of the cooling rate on synchrotron oscillation amplitude. It is shown that the numerical solution agrees well with the analytical solution at the proper limits taken by the latter. In section V, we present the numerical solution for parameters of the proof of CeC principle experiment and compare it with that obtained from macroparticle tracking in section VI. We summarize in section VII.

## II. 1-D FOKKER-PLANCK EQUATION

In the cooling section of a CeC system, a circulating ion sees a coherent energy kick induced by itself to correct its energy error as well as a random energy kick induced by its neighbours (electrons and ions). In addition, the circulating ions also get random kicks from the Intra-beam scattering (IBS). In the presence of the cooling force and the random diffusive kicks, the evolution of the longitudinal phase space density is described by the 1-D Fokker-Planck equation[8, 9]

$$\frac{\partial}{\partial t}F(I,t) - \frac{\partial}{\partial I}\left(\zeta(I) \cdot I \cdot F(I,t)\right) - \frac{\partial}{\partial I}\left(I \cdot D(I) \cdot \frac{\partial F(I,t)}{\partial I}\right) = 0 \tag{1}$$

where $F(I,t)$ is the longitudinal phase space density averaged over one synchrotron oscillation, $I$ is the amplitude of synchrotron oscillation, $\zeta(I) = -\frac{1}{I}\left\langle\frac{\Delta I_c}{T_{rev}}\right\rangle_{T_s}$ and $D(I) = \frac{1}{I}\frac{\left\langle\Delta I_d^2\right\rangle_{T_s}}{2T_{rev}}$ are the cooling rate and the diffusion coefficient averaged over one synchrotron oscillation. The diffusion coefficients are to be calculated from the summation of all random kicks. Throughout the context, we assume that the ion bunch length is much smaller than the fundamental wavelength of the RF cavities and consequently we can apply small amplitude approximation for the synchrotron oscillation. The action-angle variables under this approximation are given by[10]

$$P = \sqrt{2I}\cos w \tag{2}$$

and

$$\phi = \sqrt{2I}\sin w, \tag{3}$$

where $\phi$ is the RF phase and

$$P \equiv -h\frac{|\eta|}{\nu_s}\frac{\Delta p}{p}, \tag{4}$$

is the normalized energy deviation of the ion. The unperturbed motion of the ions can be derived from the Hamiltonian

$$H_0 = \frac{1}{2}\omega_0\nu_s P^2 + \omega_0\nu_s\frac{1}{2}\phi^2 = \omega_0\nu_s I, \tag{5}$$

and the distribution function, $F(I,t)$, satisfies the following relation

$$\int_0^\infty F(I,t)dI = \frac{N}{2\pi}, \tag{6}$$

with $N$ being the total number of ions in the bunch. It follows from eq. (2), eq. (3) and eq. (6) that

$$\int_{-\infty}^{\infty} F\left(\frac{P^2+\phi^2}{2},t\right)dPd\phi = \int_{-\infty}^{\infty} F\left(\frac{P^2+\phi^2}{2},t\right)\frac{\partial(P,\phi)}{\partial(I,w)}dIdw$$

$$= \int_{0}^{2\pi}\int_{0}^{\infty} F(I,t)dIdw \quad , \quad (7)$$

$$= N$$

and hence the distribution function in canonical variable, $(\phi,P)$, is given by

$$K(P,\phi,t) = F\left(\frac{1}{2}P^2 + \frac{1}{2}\phi^2, t\right). \quad (8)$$

At equilibrium, the first term at the L.H.S. of eq. (1) vanishes and we obtain the following solution for the equilibrium distribution of the ion beam:

$$F_{eq}(I) = A\exp\left\{-\int \frac{\zeta(I)}{D(I)}dI\right\}. \quad (9)$$

### III. ANALYTICAL SOLUTIONS IN THE LIMIT OF ZERO DIFFUSION

If the expected energy kicks from diffusion are much smaller than that from the coherent cooling kicks, i.e. $\langle D(I)\rangle \ll \langle \xi(I)I\rangle$, with the angled bracket denoting averaging over all ions, we can neglect the third term on the L.H.S. of eq. (1) and obtain

$$\frac{\partial}{\partial t}\tilde{F}(I,t) - \alpha(I)\frac{\partial}{\partial I}\tilde{F}(I,t) = 0. \quad (10)$$

with

$$\tilde{F}(I,t) \equiv \zeta(I)\cdot I \cdot F(I,t) , \quad (11)$$

and

$$\alpha(I) \equiv I \cdot \zeta(I). \quad (12)$$

Before proceeding any further, we consider a case when the coefficient $\alpha(I)$ does not depend on $I$, i.e. $\alpha(I) = \alpha_0$. For this special case, eq. (10) reduces to the wave equation

$$\frac{\partial}{\partial t}\tilde{F}_w(I,t) - \alpha_0 \frac{\partial}{\partial I}\tilde{F}_w(I,t) = 0 , \quad (13)$$

which has the general solution of the form

$$\tilde{F}_w(I,t) = G_w(I + \alpha_0 t) \ . \tag{14}$$

Inspired by above observation, we make an ansatz and assume the solution of eq. (10) for arbitrary $\alpha(I)$ satisfies

$$\tilde{F}(X(\tau), Y(\tau)) = G \tag{15}$$

where

$$\begin{cases} I = X(\tau) \\ t = Y(\tau) \end{cases} \tag{16}$$

defines a contour in the $(I,t)$ plane, $\tau$ is a parameter specifying the location along the contour and $G$ is the value that $\tilde{F}$ takes at the contour which does not depends on the location parameter $\tau$. Taking the first derivative of eq. (15) with respect to $\tau$ yields

$$\begin{aligned}
&\frac{d}{d\tau} \tilde{F}(X(\tau), Y(\tau)) \\
&= \frac{dX(\tau)}{d\tau} \cdot \frac{\partial}{\partial I} \tilde{F}(I,t) \bigg|_{\substack{I=X(\tau) \\ t=Y(\tau)}} + \frac{dY(\tau)}{d\tau} \cdot \frac{\partial}{\partial t} \tilde{F}(I,t) \bigg|_{\substack{I=X(\tau) \\ t=Y(\tau)}} \\
&= 0
\end{aligned} \tag{17}$$

Comparing eq. (17) with eq. (10) yields

$$\frac{dX(\tau)}{d\tau} = -\alpha(X(\tau)) \ , \tag{18}$$

and

$$\frac{dY(\tau)}{d\tau} = 1 \ . \tag{19}$$

Solving eq. (19) generates

$$t = \tau + t_0 \ , \tag{20}$$

and the solution of eq. (18) is

$$\tau = -\int \frac{dI}{\alpha(I)} + C \ , \tag{21}$$

where eq. (16) is applied to eq. (20) and (21). Inserting eq. (21) into eq. (20) leads to

$$C = t - t_0 + \int \frac{dI}{\alpha(I)} . \tag{22}$$

Eq. (22) defines a series of contours and each contour is specified by the value of $C$. If $(I,t)$ stays in the same contour, the value of $\tilde{F}(I,t)$ does not change. Thus, we obtain the general solution of eq. (10)

$$\tilde{F}(I,t) = G(C), \tag{23}$$

with $C$ given by eq. (22). At $t = t_0$, the solution must satisfy a given initial condition

$$\tilde{F}(I,t_0) = \tilde{F}_0(I), \tag{24}$$

and imposing the initial condition to eq. (23) leads to

$$G(h(I)) = \tilde{F}_0(I). \tag{25}$$

with

$$h(I) \equiv \int \frac{dI}{\alpha(I)} . \tag{26}$$

For any value of

$$C = h(I), \tag{27}$$

eq. (25) requires

$$G(C) = \tilde{F}_0(h^{-1}(C)) . \tag{28}$$

Inserting eq. (23) and (28) into eq. (11) yields the solution of eq. (10) which satisfy the initial condition of eq. (24)

$$F(I,t) = \frac{h^{-1}(C)\zeta(h^{-1}(C))F_0(h^{-1}(C))}{\zeta(I) \cdot I} . \tag{29}$$

Eq. (29) is a general result valid for any given form of cooling rate, $\zeta(I)$. In practice, however, a close form solution of $h^{-1}(C)$ does not always exist for the specific $\zeta(I)$ and consequently the inverse function must be found numerically. As an analytically tractable example, we assume the dependence of the cooling rate on the synchrotron oscillation action takes the following form

$$\zeta(I) = \zeta_0 \frac{I_e}{I + I_e}, \tag{30}$$

where $\zeta_0$ is the cooling rate for ions with close-to-zero synchrotron oscillation amplitude and $I_e$ is a parameter determining effective cooling range in the longitudinal direction[2]. Inserting eq. (30) and eq. (12) into eq. (26) yields

$$h(I) = \frac{1}{\zeta_0} \int \frac{1 + \frac{I}{I_e}}{I} dI = \frac{1}{\zeta_0} \ln\left(\frac{I}{I_e} \exp\left(\frac{I}{I_e}\right)\right), \tag{31}$$

and the inverse function of $h(I)$ is given by solving $I$ for equation

$$h(I) = \frac{1}{\zeta_0} \ln\left(\frac{I}{I_e} \exp\left(\frac{I}{I_e}\right)\right) = C. \tag{32}$$

The solution of eq. (32) reads

$$I = h^{-1}(C) = I_e P_{\log}\left[\exp(\zeta_0 C)\right], \tag{33}$$

where the product logarithm function, $P_{\log}(x)$, is the inverse function of

$$w(x) = xe^x, \tag{34}$$

i.e.

$$P_{\log}(x) = w^{-1}(x). \tag{35}$$

Inserting eq. (33) into eq. (29) generates

$$F(I,t) = \left(1 + \frac{I_e}{I}\right) \cdot \frac{P_{\log}\left[\exp(\zeta_0 C)\right] F_0\left(I_e P_{\log}\left[\exp(\zeta_0 C)\right]\right)}{1 + P_{\log}\left[\exp(\zeta_0 C)\right]}. \tag{36}$$

We take the initial ion distribution as

---

[2] $I_e$ is determined by the bunch length of the electron beam which can be significantly shorter than that of the ion beam (see eq. (42)). For the proof of CeC principle experiment at RHIC, the electron bunch is more than two orders of magnitude shorter than the ion bunch.

$$F_0(I) = \frac{N}{2\pi I_{ion}} \exp\left(-\frac{I}{I_{ion}}\right), \tag{37}$$

where $I_{ion}$ is a parameter determined by the longitudinal emittance of the ion bunch. Inserting eq. (26) and eq. (31) into eq. (22), and taking $t_0 = 0$ lead to

$$C = t + \frac{1}{\zeta_0} \ln\left(\frac{I}{I_e} \exp\left(\frac{I}{I_e}\right)\right). \tag{38}$$

Making use of eq. (37) and (38), eq. (36) becomes

$$F(I,t) = \frac{N}{2\pi I_{ion}} g\left(\frac{I}{I_e}\right), \tag{39}$$

with

$$g(\eta,t) = \left(1 + \frac{1}{\eta}\right) \frac{P_{\log}(\eta \exp(\zeta_0 t + \eta))}{1 + P_{\log}(\eta \exp(\zeta_0 t + \eta))} \exp\left(-\frac{I_e}{I_{ion}} P_{\log}(\eta \exp(\zeta_0 t + \eta))\right). \tag{40}$$

Using eq. (8), the line number density profile of the ion beam is then given by the following expression

$$\begin{aligned}\rho_{ion}(t,z) &= k_{rf} \int_{-\infty}^{\infty} F\left(\frac{1}{2} k_{rf}^2 z^2 + \frac{1}{2} P^2, t\right) dP \\ &= \frac{N}{2\pi} \frac{l_e}{\sigma_i^2} \int_{-\infty}^{\infty} g\left(\frac{z^2}{l_e^2} + y^2, t\right) dy\end{aligned}, \tag{41}$$

where $z$ is the longitudinal location along the ion bunch, $l_e$ is one half of electron full bunch length, $k_{rf} = 2\pi / \lambda_{rf}$ is the RF wavevector, $\lambda_{rf}$ is the RF wavelength and $\sigma_i$ is the RMS bunch length of the ion beam. In deriving eq. (41), we used

$$l_e \equiv \frac{\phi_{e,\max}}{k_{rf}} = \frac{\sqrt{2I_e}}{k_{rf}}, \tag{42}$$

with $\phi_{e,\max}$ being the maximal RF phase for an ion with synchrotron oscillation action of $I_e$. From eq. (37), we also obtained the initial RMS bunch length of ion beam as

$$\sigma_i = \frac{\sqrt{I_{ion}}}{k_{rf}} = l_e \sqrt{\frac{I_{ion}}{2I_e}}, \tag{43}$$

and the initial RMS energy spread of the ion beam as

$$\sigma_\delta = \frac{v_s}{h|\eta|}\sqrt{I_{ion}}. \tag{44}$$

As an example of applying eq. (41), Fig. 1 shows line density of an ion bunch after being cooled by a short electron bunch sitting in its center with cooling rate profile given by eq. (30). The half electron bunch length is $l_e = 0.045\sigma_i$. As shown in fig.1 (Left), a small blip appears after the ion bunch being cooled for a couple of local cooling time, $2\zeta_0^{-1}$. After the cooling time increases to $10\zeta_0^{-1}$, a dense core appears around the ion bunch center. Since the local blip appears much faster than the overall reduction of the longitudinal emittance, it could potentially be used as a diagnostic tool for optimizing cooling in commission a CeC system. However, the formation of the core is both due to the localized cooling as well as the absence of a mechanism to drive the particles out of the dense region. As we will see in the next section, the core is smoothed out for non-zero diffusion coefficient since the diffusive kicks tend to move particle out of the dense region.

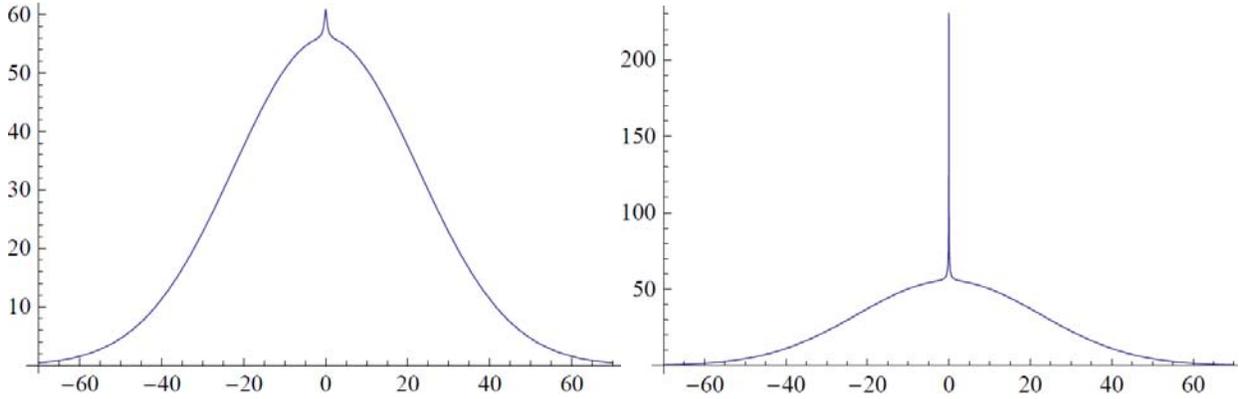

Figure 1: ion beam line density profile, $\rho_{ion}(t,z)$, as calculated from eq. (41). The abscissa is the longitudinal distance from the center of electron bunch in unit of one half of electron full bunch length, $l_e$. The ordinate is the line density of the ion bunch in unit of $\frac{N}{2\pi}\frac{l_e}{\sigma_i^2}$. (Left) ion beam profile after $t = 2\zeta_0^{-1}$; (Right) ion beam profile after $t = 10\zeta_0^{-1}$. For both plots, it is assumed $I_e / I_{ion} = 10^{-3}$.

### IV. NUMERICAL SOLUTIONS FOR FINITE DIFFUSION

In the presence of non-zero diffusion coefficient, finding analytical solution of eq. (10) is usually difficult and numerical approach is pursued. Using the following definitions of normalized variables: $\bar{D}(r) \equiv D(r^2 I_e)/D(0)$, $\bar{\zeta}(r) \equiv \zeta(r^2 I_e)/\zeta_0$, $\bar{D}_0 \equiv D(0)/(\zeta_0 I_e)$, $\bar{t} = t\zeta_0$, $R(r,\bar{t}) \equiv \frac{2\pi I_{ion}}{N} F(r^2 I_e, \bar{t}\zeta_0^{-1})$ and

$$r \equiv \sqrt{I/I_e}, \tag{45}$$

eq. (1) can be re-written as

$$r\frac{\partial R(r,\bar{t})}{\partial \bar{t}} + \alpha(r)\frac{\partial R(r,\bar{t})}{\partial r} + \beta(r)\frac{\partial^2 R(r,\bar{t})}{\partial r^2} + \gamma(r) R(r,\bar{t}) = 0, \tag{46}$$

where

$$\alpha(r) = -\frac{r^2}{2}\bar{\zeta}(r) - \frac{\bar{D}_0}{4}\bar{D}(r) - \frac{\bar{D}_0 r}{4}\frac{d\bar{D}(r)}{dr}, \tag{47}$$

$$\beta(r) = -\frac{\bar{D}_0 r}{4}\bar{D}(r), \tag{48}$$

and

$$\gamma(r) = -\frac{r^2}{2}\frac{d\bar{\zeta}(r)}{dr} - r\bar{\zeta}(r). \tag{49}$$

The difference equation derived from eq. (46) reads

$$\frac{\beta_j}{\Delta r^2} R_{j-1}^{n+1} + \left(\frac{r_j}{\Delta \bar{t}} - \frac{\alpha_j}{\Delta r} - 2\frac{\beta_j}{\Delta r^2} + \gamma_j\right) R_j^{n+1} + \left(\frac{\alpha_j}{\Delta r} + \frac{\beta_j}{\Delta r^2}\right) R_{j+1}^{n+1} = \frac{r_j}{\Delta \bar{t}} R_j^n \tag{50}$$

for $2 \leq j < N$ with $N$ being the index of the last bin in the grid of $r$,

$$-\frac{\alpha_j}{\Delta r} R_1^{n+1} + \frac{\alpha_j}{\Delta r} R_2^{n+1} = 0 \tag{51}$$

for $j=1$, and

$$\frac{\beta_N}{\Delta r^2} R_{N-1}^{n+1} + \left(\gamma_N + \frac{r_N}{\Delta t} - \frac{\alpha_N}{\Delta r} - 2\frac{\beta_N}{\Delta r^2}\right) R_N^{n+1} = \frac{r_N}{\Delta \bar{t}} R_N^n \tag{52}$$

for $j = N$. In deriving eq. (51) and (52), the following boundary conditions are applied:

$$\left.\frac{\partial R(r,\bar{t})}{\partial r}\right|_{r=0} = 0 , \tag{53}$$

and

$$R(r,\bar{t})\big|_{r\to\infty} = 0. \tag{54}$$

Numerical solution of eq. (50)-(52) is obtained by applying the subroutine, TRIDAG, from Numerical Recipes[11]. After obtaining the phase space density, $R(r,\bar{t})$, the line number density of the ion bunch is given by

$$\rho(z,t) = \frac{N}{2\pi} \frac{l_e}{\sigma_i^2} \int_{-\infty}^{\infty} R\left(\sqrt{y^2 + \frac{z^2}{l_e^2}}, t\zeta_0\right) dy . \tag{55}$$

Fig.2 shows the ion bunch line density profile after $3\zeta_0^{-1}$ of cooling where the initial ion distribution is given by eq. (37), i.e.

$$R_0(r) = \exp(-r^2 / r_0^2) \tag{56}$$

with

$$r_0^2 \equiv I_{ion} / I_e , \tag{57}$$

and to compare with analytical result in eq. (41), we take

$$\bar{\zeta}(r) = \frac{1}{1+r^2} . \tag{58}$$

The numerical method is applicable to arbitrary dependence of diffusion coefficient on synchrotron oscillation amplitude. In fig. 2 (magenta dots), to illustrate the effects of diffusion, we take

$$D(r^2 I_e) = \frac{\bar{D}_0}{1+r^2} I_e \zeta_0 , \tag{59}$$

with $\bar{D}_0 = 100$ as an example. As shown in Fig.2, the numerical solution (blue dashes) reproduces the analytical result (red solid) in the absence of diffusion. For non-vanishing diffusion coefficient (magenta dots), however, the central blip is smoothed out. As we will see in the next section, the smoothing effects due to diffusion does not rely on the specific form of cooling rate and diffusion coefficient, i.e. eq. (58), and similar influences of diffusion to the ion bunch profile are observed for a more realistic cooling profile derived for the FEL-based CeC simulation.

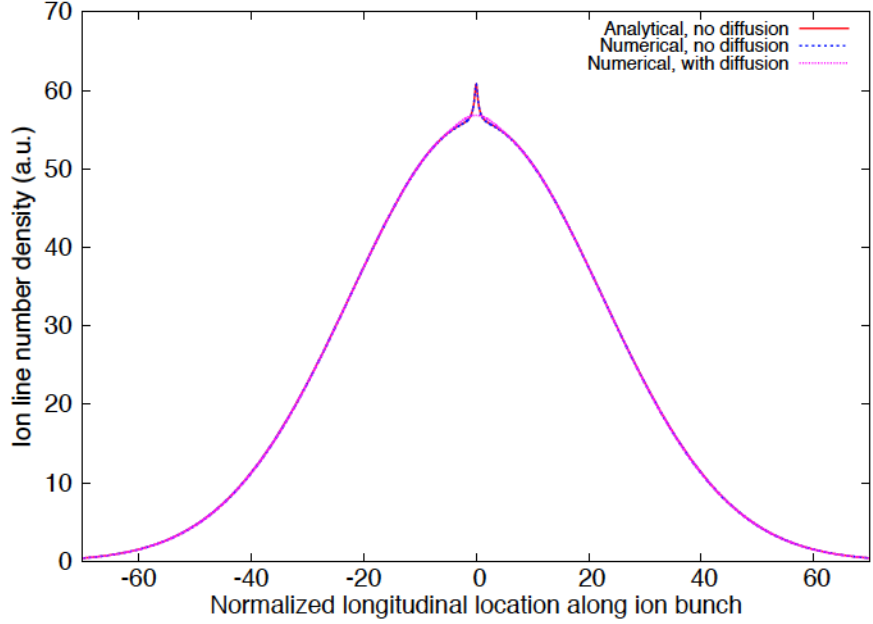

Figure 2: Comparing the numerical solution of eq. (50)-(52) (blue) with the analytical solution as calculated from eq. (39) (red). The abscissa is the normalized longitudinal location in unit of $l_e$ and the ordinate is the ion bunch line number density in unit of $\dfrac{N}{2\pi}\dfrac{l_e}{\sigma_i^2}$ as calculated from eq. (55). The red and blue curve show results for zero diffusion. The magenta curve shows the numerical solution of eq. (50)-(52) with diffusion coefficient, $\bar{D}_0 = 100$, and the profile $\bar{D}(r) = 1/(1+r^2)$. All curves are plotted for $\bar{t} = 2$ and $r_0^2 = I_{ion}/I_e = 1000$.

## V. EVALUATION FOR FEL-BASED COHERENT ELECTRON COOLING

The assumed dependence of cooling rate on synchrotron oscillation amplitude in eq. (30) is convenient for obtaining analytical solution, eq. (39), as well as validating the numerical method of solving eq. (50)-(52). In practice, however, the dependences of both the cooling rate and diffusion coefficient on synchrotron amplitude, after averaged over one synchrotron oscillation period, should be determined by the specific cooling scheme such as the longitudinal profile of the electron bunch, whether the electron bunch is painted around the ion bunch, and how cooling force and diffusive kick depend on local electron properties. The single pass energy kick received by an ion as it travels through the storage ring is given by

$$\Delta \delta\gamma_{j,n} = \delta\gamma_{j,n+1} - \delta\gamma_{j,n} = -g_\gamma \sin\left(\dfrac{2\pi R_{56}}{\lambda_0 \gamma_0}\delta\gamma_{j,n}\right) + d_{ion}X_{j,n} + d_e Y_{j,n} + d_{IBS}Z_{j,n}\,, \qquad (60)$$

where

$$\delta\gamma_{j,n} = \frac{v_s\gamma_0}{h|\eta|} P_{j,n} \tag{61}$$

is the energy deviation (normalized to its rest energy, $A_i m_u c^2$) of the $j^{th}$ ion in its $n^{th}$ circulation around the ring,

$$g_\gamma \equiv \frac{Z_i e E_p l}{A_i m_u c^2} \tag{62}$$

is the amplitude of the self-induced coherent energy kick received by the ion in the cooling section, $\lambda_0$ is the optical wavelength of the FEL amplifier, $f_{rev}$ is the revolution frequency of the ions and $R_{56}$ is the longitudinal dispersion from the modulator to the kicker of the cooling section. For ions with $\frac{2\pi R_{56}}{\lambda_0\gamma_0}\delta\gamma_j \ll 1$, eq. (60) can be written as

$$\Delta\delta\gamma_{j,n} = -g_\gamma \frac{2\pi R_{56}}{\lambda_0\gamma_0}\delta\gamma_{j,n} + d_{ion}X_{j,n} + d_e Y_{j,n} + d_{IBS}Z_{j,n} . \tag{63}$$

The local cooling time, $T_0$, and its inversion, the local cooling rate, $\zeta_0$, are related to the coherent energy kick, $g_\gamma$, by

$$T_0 = \zeta_0^{-1} = \frac{1}{f_{rev}} \cdot \frac{\lambda_0\gamma_0}{g_\gamma 2\pi R_{56}} . \tag{64}$$

For each circulation around the ring, the reduction of the ion's longitudinal oscillation action due to cooling is given by

$$\Delta I_c = \frac{1}{2}\Delta(P^2+\phi^2) = P\Delta P_c , \tag{65}$$

where

$$\Delta P_c = \frac{h|\eta|}{v_s\gamma}\Delta\delta\gamma_c . \tag{66}$$

Inserting

$$\Delta\delta\gamma_c = -g_\gamma \frac{2\pi R_{56}}{\lambda_0\gamma_0}\delta\gamma = -\zeta_0 T_{rev}\delta\gamma , \tag{67}$$

into eq. (65) and using eq. (61) give

$$\Delta I_c = -\zeta_0 T_{rev} P^2 = -2I\zeta_0 T_{rev} \cos^2 w, \qquad (68)$$

where $T_{rev}$ is the revolution period of the ion and we use eq. (2) in deriving the second equation of (68). Averaging eq. (68) over one synchrotron oscillation leads to

$$\zeta(r) = -\frac{1}{I}\left\langle \frac{\Delta I_c}{T_{rev}} \right\rangle_{T_s} = \zeta_0 \bar{\zeta}(r), \qquad (69)$$

where $r = \sqrt{I/I_e}$, $I_e = \frac{1}{2}(k_{rf} l_e)^2$,

$$\bar{\zeta}(r) = \begin{cases} \dfrac{2}{\pi}\arcsin\left(\dfrac{1}{r}\right) + \dfrac{2}{\pi r^2}\sqrt{r^2-1}; & \text{for } r \geq 1 \\ 1 & ; \quad \text{for } r < 1 \end{cases}, \qquad (70)$$

and for $r > 1$, we used the following relation to derive eq. (70)

$$\left\langle \cos^2 w \right\rangle_{T_s} = \frac{2}{2\pi}\int_{-\theta}^{\theta} \cos^2 w\, dw = \frac{\theta}{\pi} + \frac{1}{\pi}\sin\theta\cos\theta, \qquad (71)$$

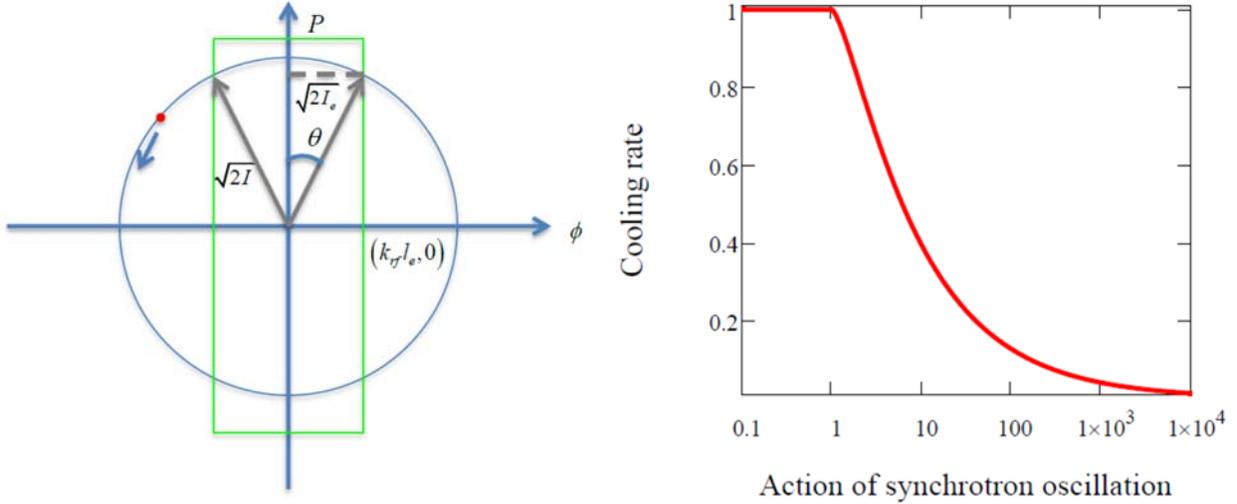

Figure 3: Averaged cooling rate over a synchrotron oscillation period. (Left) Illustration of an ion passing through the cooling electron beam while conducting synchrotron oscillation in its longitudinal phase space with the green box representing the longitudinal range of electron bunch and the red dot representing the ion; (Right) averaged cooling rate as calculated from eq. (69) in unit of $\zeta_0$ (see eq. (64)). The ordinate in the right plot is the action of the ion's synchrotron oscillation in unit of $I_e = k_{rf}^2 l_e^2 / 2$.

with $\sin\theta = \sqrt{I_e/I} = 1/r$ as shown in fig. 3 (Left). Fig. 3 (right) shows how the cooling rate depends on the synchrotron oscillation action of the particle as calculated from eq. (70).

The second and the third term in the R.H.S. of eq. (60) represent the diffusive kicks received by the ion in the cooling section (APPENDIX A), where $X_{j,n}$ and $Y_{j,n}$ are random numbers uniformly distributed from -1 to 1,

$$d_{ion} = g_\gamma \sqrt{\frac{3}{2}\sqrt{\pi}\rho_{ion}(z_j)}\sigma_w \approx g_\gamma \sqrt{\frac{3}{2}\sqrt{\pi}\rho_{ion}(z_e)}\sigma_w \tag{72}$$

is the amplitude of the incoherent energy kick induced by all neighbor ions and

$$d_e = \frac{g_\gamma}{Z_i}\sqrt{\frac{3}{2}\sqrt{\pi}\rho_e(z_j)}\sigma_{z,rms} \approx \frac{g_\gamma}{Z_i}\sqrt{\frac{3}{2}\sqrt{\pi}\rho_e}\sigma_w \tag{73}$$

is the amplitude of the incoherent energy kick induced by all neighbor electrons. In deriving the second equation of eqs. (72) and (73), we assume that the center of the electron bunch is located at $z_e$, the ion beam line density does not vary significantly over the range of the cooling electron beam, and the electron beam has uniform line density. The last term of the R.H.S. of eq. (60) is responsible for the accumulated energy kicks due to Intra-beam scattering (IBS) while ions traveling through the ring. $Z_{j,n}$ is a random number uniformly distributed from -1 to 1 and the kick strength, $d_{IBS}$, is to be determined by the IBS growth rate as obtained from Piwinski's formulas [12] (APPENDIX B).

For an ion with synchrotron oscillation action, $I$, a random energy kick in its $i^{th}$ circulation, $d \cdot X_i$, corresponds to a change of its longitudinal action of

$$\Delta I_d = P\Delta P_d = h\frac{|\eta|}{v_s}\frac{d}{\gamma_0}X\sqrt{2I}\cos w. \tag{74}$$

After averaging over the random variable, $X$, and the synchrotron phase, $w$, the RMS variation of $\Delta I$ is

$$\langle \Delta I_d^2 \rangle = \frac{1}{3}\left(h\frac{|\eta|}{v_s}\right)^2\left(\frac{d}{\gamma_0}\right)^2 2I\langle \cos^2 w\rangle_{T_s}. \tag{75}$$

Using eq. (71), the diffusion coefficient in the Fokker-Planck equation, eq. (1), is given by [13]

$$D(I) = \frac{\langle \Delta I_d^2 \rangle}{2T_{rev}I} = D(0)\bar{D}(I), \tag{76}$$

with

$$D(0) = \frac{d^2}{6T_{rev}\gamma^2}\left(h\frac{|\eta|}{v_s}\right)^2. \tag{77}$$

The normalized diffusion coefficient, $\bar{D}_0 \equiv D(0)/(I_e\zeta_0)$, is

$$\bar{D}_0 = \frac{d^2}{6T_{rev}\zeta_0 I_e\gamma^2}\left(h\frac{|\eta|}{v_s}\right)^2 = \frac{d^2}{3T_{rev}\zeta_0}\left(\frac{\sigma_i}{l_e\gamma\sigma_\delta}\right)^2, \tag{78}$$

where in deriving eq. (78), we used eqs. (43) and (44), i.e.

$$I_e = \frac{l_e^2}{2\sigma_i^2}I_{ion} = \frac{l_e^2}{2\sigma_i^2}\left(h\frac{|\eta|}{v_s}\sigma_\delta\right)^2. \tag{79}$$

Combining eqs. (71), (75), (76) and (78), we obtain the diffusion coefficient for the incoherent kicks along the cooling section

$$D_{cec}(r) = \zeta_0 I_e \bar{D}_{0,cec}\bar{D}_{cec}(r) \tag{80}$$

with

$$\bar{D}_{0,cec} = \frac{d_e^2 + d_{ion}^2}{3T_{rev}\zeta_0}\left(\frac{\sigma_i}{l_e\gamma\sigma_\delta}\right)^2, \tag{81}$$

and

$$\bar{D}(r) = \begin{cases} \frac{2}{\pi}\arcsin\left(\frac{1}{r}\right) + \frac{2}{\pi r^2}\sqrt{r^2-1}; & \text{for } r \geq 1 \\ 1 & ; \text{ for } r < 1 \end{cases}. \tag{82}$$

For the next step, we calculate the diffusion coefficient due to IBS. Similar to eq. (74), the increase of longitudinal action due to IBS kick is

$$\Delta I_{IBS} = P\Delta P_{ibs} = h\frac{|\eta|}{v_s}\frac{d_{IBS}(\phi)}{\gamma_0}X\sqrt{2I}\cos w, \tag{83}$$

where $d_{IBS}(\phi)$ is the IBS kick strength which is proportional to the square root of local ion density. Making use of eqs. (37) and (43), we obtain

$$d_{IBS}(\phi) = d_{0,IBS} \exp\left(-\frac{\phi^2}{4k_{rf}^2 \sigma_i^2}\right) = d_{0,IBS} \exp\left(\frac{-I}{2I_{ion}} \sin^2 w\right), \tag{84}$$

with $d_{0,IBS}$ being the kick strength at the bunch center. The variance of the action kick is obtained from eq. (83):

$$\left\langle \Delta I_{IBS}^2 \right\rangle = \frac{2I d_{0,IBS}^2}{3} \left(h\frac{|\eta|}{v_s}\right)^2 \left(\frac{1}{\gamma_0}\right)^2 \left\langle \exp\left(\frac{-I}{I_{ion}} \sin^2 w\right) \cos^2 w \right\rangle_{T_s}. \tag{85}$$

With the help from eq. (76) and the following relation

$$\left\langle \exp(-z \sin^2 w) \cos^2 w \right\rangle_{T_s} = \frac{1}{2\pi} \int_0^{2\pi} \exp(-z \sin^2 w) \cos^2 w \, dw = \frac{1}{2} e^{-\frac{z}{2}} \left[ I_0\left(\frac{z}{2}\right) + I_1\left(\frac{z}{2}\right) \right], \tag{86}$$

we obtain

$$D_{IBS}(I) = \frac{\left\langle \Delta I_{IBS}^2 \right\rangle}{2T_{rev} I} = I_e \zeta_0 \bar{D}_{0,IBS} \bar{D}_{IBS}(I), \tag{87}$$

where $I_0(x)$ and $I_1(x)$ are modified Bessel function,

$$\bar{D}_{0,IBS} = \frac{d_{0,IBS}^2}{3T_{rev} \zeta_0} \left(\frac{\sigma_i}{l_e \gamma \sigma_\delta}\right)^2, \tag{88}$$

and

$$\bar{D}_{IBS}(I) = e^{-\frac{I}{2I_{ion}}} \left[ I_0\left(\frac{I}{2I_{ion}}\right) + I_1\left(\frac{I}{2I_{ion}}\right) \right]. \tag{89}$$

Inserting eqs. (45) and (57) into eq. (87) yields

$$D_{IBS}(r) = I_e \zeta_0 \bar{D}_{0,IBS} \bar{D}_{IBS}(r), \tag{90}$$

with

$$\bar{D}_{IBS}(r) = e^{-\frac{r^2}{2r_0^2}} \left[ I_0\left(\frac{r^2}{2r_0^2}\right) + I_1\left(\frac{r^2}{2r_0^2}\right) \right]. \tag{91}$$

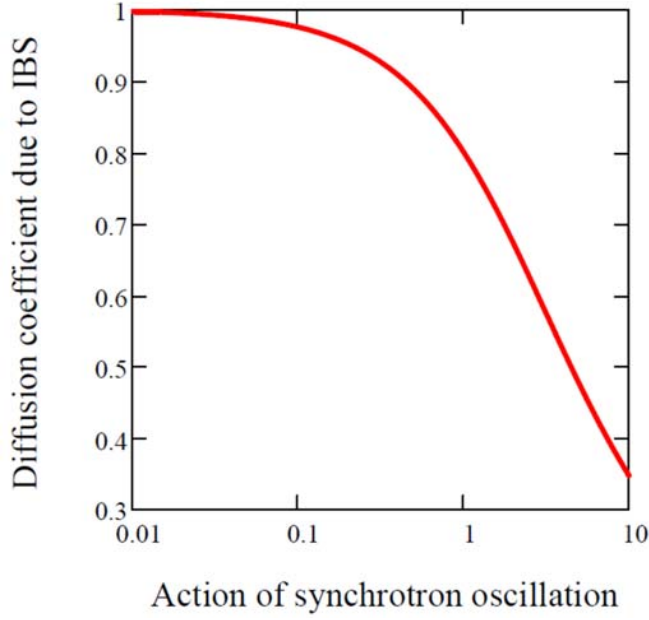

Figure 4: Dependence of IBS induced diffusion coefficient on the synchrotron oscillation action of an ion. The ordinate is the normalized diffusion coefficient induced by IBS as calculated from eq. (89) and the abscissa is the synchrotron oscillation action of an ion in unit of $I_{ion}=k_{rf}^2\sigma_i^2$.

| Electron beam parameters | | Ion beam parameters | |
|---|---|---|---|
| Peak current, A | 40 | Charge number, $Z_{ion}$ | 79 |
| Full bunch length, ps | 25 | Bunch intensity | $10^8$ |
| Norm. emittance, RMS, μm | 5 | Bunch length, RMS, ns | 3.06 |
| Relative energy spread, RMS | $10^{-3}$ | Relative energy spread, RMS | $3.35\times10^{-4}$ |
| Beam energy, $\gamma$ | 28.66 | RF frequency, MHz | 28 |

Table 1: Beam parameters for the proof of CeC principle experiment

| Gain of FEL amplification | 80 | FEL wavelength, $\mu m$ | 30.5 |
|---|---|---|---|
| Peak correcting field, V/m | 36 | $R_{56}$, cm | 1.2 |
| Kicker length, m | 3 | Coherent length, $\sigma_w$, mm | 0.54 |
| Coherent kick amplitude, $g_\gamma$ | $4.657\times10^{-8}$ | Local cooling time ($T_0$), s | 3.185 |
| CeC diffusive kick from neighbor ions, $d_{ion}$ | $1.163\times10^{-5}$ | CeC diffusive kick from electrons, $d_e$ | $2.038\times10^{-5}$ |
| IBS diffusive kick at bunch center, $d_{IBS}(0)$ | $1.886\times10^{-6}$ | | |

Table 2: CeC system parameters of the proof of CeC principle experiment

The total diffusion coefficient is obtained by the summation of eqs. (80) and (90)

$$D(r) = D_{cec}(r) + D_{IBS}(r). \tag{92}$$

For the parameters of the proof of CeC principle experiment listed in table 1 and table 2, the normalized diffusion coefficients are $\bar{D}_{0,cec} = 2.97 \times 10^4$ and $\bar{D}_{0,IBS} = 192$ respectively. Fig. 4 (right) shows the ion bunch current profile around the bunch center after ~1 minute of cooling with the nominal (Green) and artificially reduced diffusion coefficients (Blue, Magenta and Black). As shown in fig. 5 (right), the local blip is fully suppressed for the nominal parameters and starts to be visible if the normalized diffusion coefficient is reduced by two orders of magnitudes or more. According to eqs. (72), (73) and (81), the normalized diffusion coefficients due to CeC, $\bar{D}_{0,cec}$, decreases linearly with $g_\gamma$ and one way to reduce diffusion in the CeC process is to reduce self-induced energy kick, $g_\gamma$. However, the local cooling time, $T_0$, increases proportionally with $g_\gamma^{-1}$ and hence the normalized maximal diffusion coefficient due to IBS, $\bar{D}_{0,IBS}$, increases linearly with $g_\gamma^{-1}$. In addition, increasing $T_0$ will also make the overall process slower, defeating the purpose of using the blip as a fast-diagnostic signal.

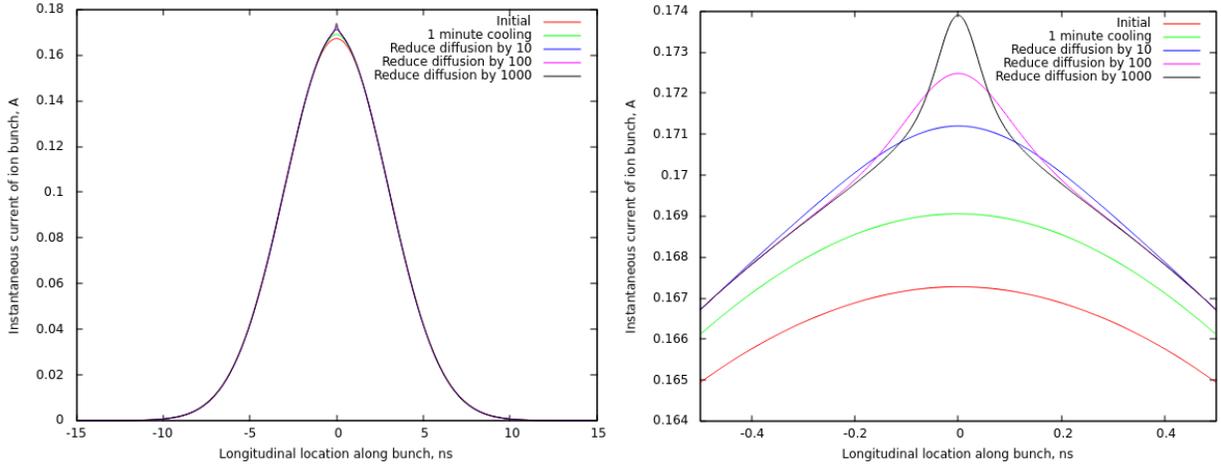

Figure 5: Ion bunch current profile after 1 minutes of cooling and its dependence on diffusion coefficient. These plots are generated by solving eq. (50)-(52) with parameters listed in Tables 1 and 2. The cooling rate and diffusion coefficients are calculated from eqs. (69) and (92). The red curve shows the initial ion bunch current profile and the green curve shows the ion bunch profile after 1 minute of cooling. The blue, magenta and black curve shows the ion bunch profiles after 1 minute of cooling with diffusion coefficient reduced by a factor of 10, 100 and 1000 from the nominal values (green). The left plot shows the overall bunch profile and the right plot shows the zoomed-in region around the bunch center.

## VI. MACRO-PARTICLE TRACKING

The numerical method of solving Fokker-Planck equation as developed in section IV has the advantages of requiring minimal computational time (a few minutes in a pc) and resources. However, its applicability is limited to the scheme where the cooling rate and diffusion coefficient do not change significantly throughout the process, the ion bunch length is small so that the Hamiltonian of eq. (5) is accurate and the cooling force is proportional to the energy deviation of the particle. To cross-check the results obtained in section V as well as to provide a more versatile tool for general cooling scheme, we have developed a macro-particle tracking code. As shown in fig. 6, typically 0.2~6 millions of macro-ions are generated when the simulation starts. The longitudinal coordinates of each macro-ion are then updated according to the rf voltage it sees and the phase slip factor of the lattice:

$$\bar{\varepsilon} = \varepsilon + \frac{q}{mc^2} V_{rf}(\tau), \qquad (93)$$

and

$$\bar{\tau} = \tau + \frac{T_{rev}\eta}{\beta^2 \gamma_0} \bar{\varepsilon}, \qquad (94)$$

where $q$ is the charge of the ion, $m$ is the mass of the ion, $c$ is speed of light, $\varepsilon$ is the energy deviation of the macro-ion in unit of $mc^2$, $\tau$ is the arriving time of the macro-ion, $V_{rf}(\tau)$ is the rf voltage seen by an ion, $\gamma_0 mc^2$ is the energy of the reference ion, $T_{rev}$ is the revolution period and $\eta$ is the phase slip factor. The update of the transverse coordinate uses one turn linear transfer matrix:

$$\begin{pmatrix} \bar{x} \\ \bar{p}_x \\ \bar{y} \\ \bar{p}_y \end{pmatrix} = \begin{pmatrix} \cos\psi_x & \sin\psi_x & 0 & 0 \\ -\sin\psi_x & \cos\psi_x & 0 & 0 \\ 0 & 0 & \cos\psi_y & \sin\psi_y \\ 0 & 0 & -\sin\psi_y & \cos\psi_y \end{pmatrix} \begin{pmatrix} x \\ p_x \\ y \\ p_y \end{pmatrix}, \qquad (95)$$

where $\psi_x$ and $\psi_y$ are the one turn phase advances of horizontal and vertical betatron motion.

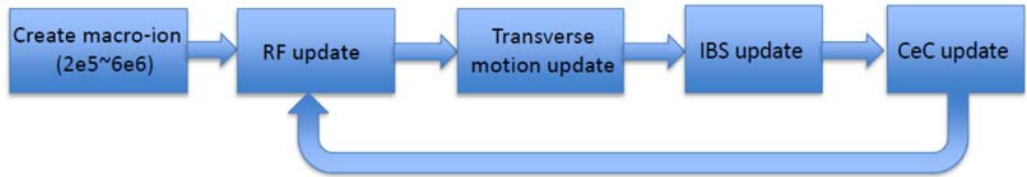

Figure 6: Flowchart illustration of the tracking process.

A random 3-D Langevin kick is applied to each macro-ion every turn to account for effects due to IBS. The R.M.S. amplitude of the kick is determined by the growth rate as calculated from the Piwinski's formula (APPENDIX B). Local ion line density is used in the IBS growth rate calculations.

To implement the one-turn update due to CeC, we first estimate how the ions are mixed from turn to turn by synchrotron oscillation. The synchrotron period for the CeC experiment is about 4000 revolutions. With the RMS ion bunch length of 3 ns, the average longitudinal slippage of a typical ion in one revolution is

$$\langle \Delta\tau_{1\sigma} \rangle \sim \frac{4 \cdot 3 ns \cdot c}{4000} = 0.9 mm,$$

which is ~30 times larger than the optical wavelength of the FEL amplifier (30.5 μm). Consequently, no phase information is preserved after one revolution and the incoherent kicks due to neighbor ions (and cooling electrons) can be implemented as a random Langevin kick as shown in eq. (60).

As shown in Fig. 7, the prediction from solving Fokker-Planck equation (red-dash) agrees very well with that obtained from macro-particle tracking if linear cooling force is adopted in the tracking (magenta). It is worth noting that the peak current from solving Fokker-Planck equation (red-dash) is about 1~2% higher than that obtained from tracking with linear cooling force (magenta), which is likely due to the static diffusion coefficients assumed in the Fokker-Planck equation, while the tracking algorithm calculates diffusive kick from the updated bunch profile. Fig. 7 also shows that the cooling effect is less pronounced if more realistic Sinusoidal cooling force is applied in the tracking, which is to be expected as ions with large synchrotron amplitude get reduced cooling or even anti-cooling during the process.

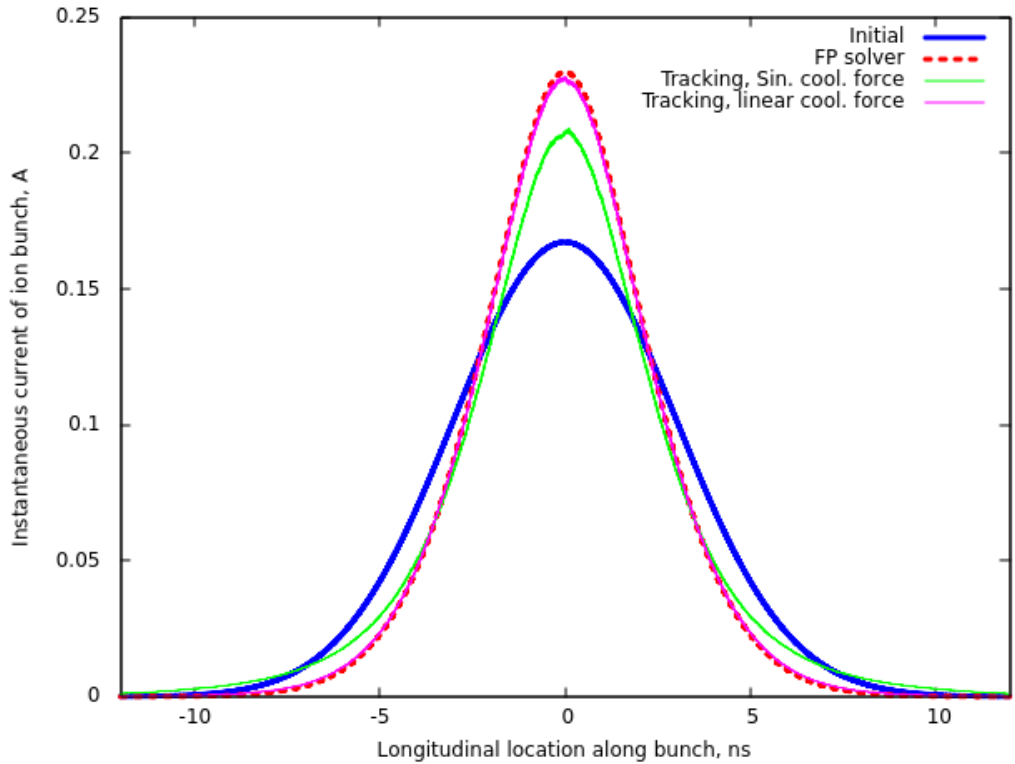

Figure 7: The ion bunch current profiles after 40 minutes of cooling as obtained by solving eqs. (50)-(52) (red) and through macro-particle tracking (green and magenta). The green curve shows the tracking results with sinusoidal cooling force as described by eq. (60), while the magenta curve shows the tracking results with linear cooling force given by eq. (63).

The technique described above are also used to study the tolerance of a CeC system to the noise level in the electron beam. Comparing the results from macro-particle tracking with that from numerically solving Fokker-Planck equation is illustrative for the limitation of the latter. Fig. 8 shows the current profiles of an ion bunch after 40 minutes of cooling by electrons with various noise levels. Parameters in tables 1 and 2 are used in generating fig. 8. Fig. 8 (a) is obtained by numerically solving eq. (50)-(52) and fig. 8 (b) is the results of macro-particle tracking. While results from both approaches predict that heating due to diffusion dominates the CeC process once the noise in the electron beam is more than a factor of 3 higher than its natural noise level, the peak currents of cooled bunch in fig. 8 (b) are all lower than that shown in fig. 8 (a) due to the approximation of linear cooling force applied in the Fokker-Planck solver. On the other hand, the witness bunch (the black curve) in fig. 8 (b) has slightly higher peak current than that in fig. 8 (a) since the IBS rate decreases with the peak current but in Fokker-Planck equation, the diffusion coefficient does not vary with time. Moreover, the number of particles is conserved in fig. 8 (a) as a result of using small amplitude Hamiltonian, i.e. eq. (5), in the Fokker-Planck equation. In tracking, particles are lost once they move out of the RF bucket and consequently bunch intensity reduces once the bunch length becomes comparable with the RF wavelength as shown in fig. 8(b).

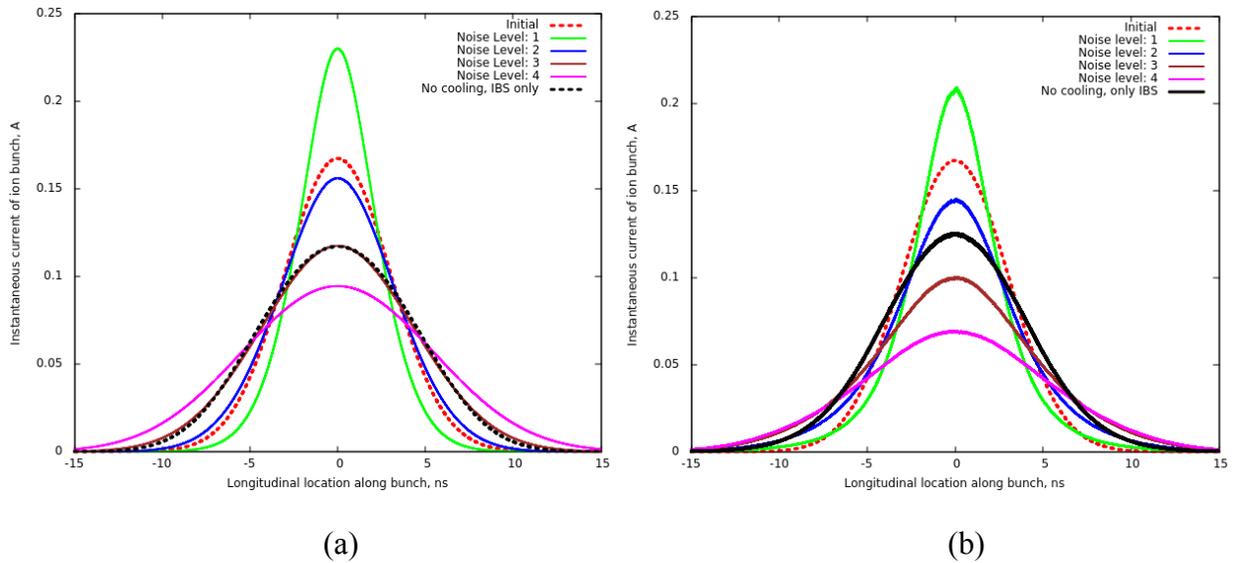

(a)          (b)

Figure 8: Tolerance of the proof of CeC principle experiment on noise level in the electrons. (a) The plots are obtained by solving eqs. (50)-(52); (b) the plots are obtained by macroparticle tracking. The red-dash curve is the initial ion bunch current profile and the black curve is the current profile of a witness bunch, i.e. an ion bunch not overlapping with electron bunch, after 40 minutes of storage. The green curve is the ion bunch profile after 40 minutes of cooling by electrons with natural noise level ($d_e = 2.038 \times 10^{-5}$ from table 1 is used in the calculation). To account for excessive noise in the electrons, increased values of $d_e$ are used in generating blue ($d_e = 4.076 \times 10^{-5}$), brown ($d_e = 6.114 \times 10^{-5}$) and magenta ($d_e = 8.152 \times 10^{-5}$) curves.

## VII. DISCUSSION AND SUMMARY

In summary, we have developed a set of tools to predict the ion bunch profile in the presence of longitudinal coherent electron cooling, which include an analytical expression for vanishing diffusion coefficient, a numerical algorithm for solving Fokker-Planck equation with linear cooling force and arbitrary dependence of cooling rate and diffusion coefficient on synchrotron oscillation action, and a macro-particle tracking code for arbitrary cooling force. In their applicable regime, we have found good agreement among these tools and used them to predict the performance of the CeC system in the proof of CeC principle experiment.

One of the insights achieved from these studies is that, for the proof of the CeC principle experiment, the ion bunch profile will not show any local blip during the cooling process due to IBS and diffusion induced by neighbor particles in the cooling section. Since any stochastic cooling mechanism, including CeC, inevitably introduces diffusive kicks due to neighbor ions, it limits the feasibility to use the high-frequency components in the beam current profile as a diagnostic tool for optimizing cooling.

## ACKNOWLEDGMENTS

The work benefits greatly from discussions with V.N. Litvinenko and M. Blaskiewicz. The macro-particle tracking code is based on various subroutines written by M. Blaskiewicz. This work was supported by Brookhaven Science Associates, LLC under Contract No. DE-AC02-98CH10886 with the U.S. Department of Energy.

# APPENDIX A. DERIVATION OF SINGLE PASS ENERGY KICK FOR FEL-BASED COHERENT ELECTRON COOLING

Using the 1-D FEL theory with high gain approximation [1], the electric field induced by a single ion at the entrance of the kicker section is

$$E_{1D}(z) = E_p \exp\left(-\frac{z^2}{2\sigma_w^2}\right)\sin\left(k_0 z - k_2^2 z^2 - \varphi_0\right), \tag{A1}$$

where $z$ is the longitudinal location with respect to the peak of the electron density wave-packet induced by the ion, $E_p = G \cdot E_0 = G \cdot Ze/(2\varepsilon_0 S)$ is the maximal electric field induced by the electron density wave-packet, $G$ is the gain of the longitudinal electric field due to FEl amplification, $S$ is the transverse area of the electron beam, $\sigma_w$ is the RMS width of the electron density wave-packet, $\varphi_0$ is a constant phase shift determined by the length of the FEL amplifier, and $k_2^2 z^2$ represents a slow phase variation along the wave-packet with $k_2^2 z^2 \sim (z/\sigma_{z,rms})^2 \ll k_0 z$. The field observed by the $j^{th}$ ion due to the wave-packet induced by the $i^{th}$ ion is given by

$$E_{1D}(z_j - \varsigma_i) = E_p \exp\left[-\frac{(z_j - \varsigma_i)^2}{2\sigma_w^2}\right]\sin\left(k_0(z_j - \varsigma_i) - k_2^2(z_j - \varsigma_i)^2 - \varphi_0\right), \tag{A2}$$

where $z_j$ is the location of the $j^{th}$ ion at the kicker section and $\varsigma_i$ is the location of the peak of the wave-packet induced by the $i^{th}$ ion. By properly delaying the ions, the ion can be placed at

$$z_j = R_{56} \cdot \delta_j + \Delta z_{sh} + \varsigma_j, \tag{A3}$$

where $\delta_j$ is the relative energy deviation of the $j^{th}$ ion, $R_{56}$ is the longitudinal dispersion from the CeC modulator section to the kicker section and $\Delta z_{sh} \in (-\pi/k_0, \pi/k_0)$ is a small delay of the electrons introduced by the phase shifter so that for an ion with zero energy deviation, the phase of the sinusoidal function in eq. (A2) with $i = j$ is $-\pi$, i.e.

$$\Delta z_{sh} \approx \left[\mod(\varphi_0, 2\pi) - \pi\right]/k_0. \tag{A4}$$

Inserting eqs. (A3) and (A4) into eq. (A2) and assuming the electric field do not change significantly inside the kicker of length $l$, we obtain the one turn energy kick received by the $j^{th}$ ion in the CeC section:

$$\Delta E_j = \Delta E_{coh,j} + \Delta E_{inc,j}, \tag{A5}$$

where

$$\Delta E_{coh,j} \equiv -Z_i e E_p l \sin(k_0 R_{56} \cdot \delta_j), \quad (A6)$$

is the energy kick induced by the $j^{th}$ ion itself, i.e. the coherent cooling kick, and the second term,

$$\Delta E_{inc,j} \equiv -Z_i e E_p l \sum_{i \neq j} \exp\left(-\frac{(\varsigma_j - \varsigma_i)^2}{2\sigma_w^2}\right) \sin\left(k_0 (R_{56}\delta_j + \varsigma_j - \varsigma_i) - k_2^2(\varsigma_j - \varsigma_i)^2\right), \quad (A7)$$

is the incoherent diffusive energy kick induced by all other ions. The variance of the incoherent kick is calculated as

$$\langle \Delta E_{inc,j}^2 \rangle = \frac{(Z_i e E_p l)^2}{2} \left\langle \sum_{i \neq j} \exp\left(-\frac{(\varsigma_i - \varsigma_j)^2}{\sigma_w^2}\right) \right\rangle, \quad (A8)$$

where the angled bracket, $\langle ... \rangle$, represents ensemble average. Assuming the ion density does not vary significantly over the width of the wave-packet, $\sigma_w$, eq. (11) reduces to

$$\langle \Delta E_{inc,j}^2 \rangle = \frac{(Z_i e E_p l)^2}{2} \int_{-\infty}^{\infty} \rho_{ion}(\varsigma_i) \exp\left(-\frac{(\varsigma_i - \varsigma_j)^2}{\sigma_w^2}\right) d\varsigma_i$$

$$\approx \frac{(Z_i e E_p l)^2}{2} \sqrt{\pi} \rho_{ion}(z_j) \sigma_w \quad (A9)$$

where $\rho_{ion}(z_j)$ is the local line number density of ions around the $j^{th}$ ion. The energy kicks in eq. (A5) can be approximated as the summation of the coherent kick as described in eq. (A6) and a random kick with the R.M.S. amplitude described by eq. (A9):

$$\Delta E_{j,N} \approx -Z_i e E_p l \sin(k_0 R_{56} \cdot \delta_j) + \sqrt{\frac{\langle \Delta E_{inc,j}^2 \rangle}{\langle X^2 \rangle}} \cdot X_{j,N}, \quad (A10)$$

where $X_{j,N}$ is a random number determining the incoherent kick acting on the $j^{th}$ ion at the $N^{th}$ turn and $\langle X^2 \rangle$ is the variance of $X_{j,N}$. For instance, if $X_{j,N}$ is a uniformly distributed random number from -1 to 1, its variance is

$$\langle X^2 \rangle = \frac{1}{2} \int_{-1}^{1} X^2 dX = \frac{1}{3}, \quad (A11)$$

and eq. (A10) becomes

$$\Delta E_{j,N} \approx -Z_i e E_p l \sin(k_0 R_{56} \cdot \delta_j) + Z_i e E_p l \sqrt{\frac{3}{2}\sqrt{\pi}\rho_{ion}(z_j)\sigma_w} \cdot X_{j,N} . \quad (A12)$$

Following a similar derivation, the incoherent kick due to shot noise from cooling electrons is derived as

$$\Delta E_{j,N}^e \approx e E_p l \sqrt{\frac{3}{2}\sqrt{\pi}\rho_e(z_j)\sigma_w} \cdot X_{j,N} , \quad (A13)$$

where $\rho_e(z_j)$ is the line number density of the electrons at location $z_j$.

## APPENDIX B. ENEERGY KICK DUE TO INTRA-BEAM SCATTERING [14]

Here, we try to derive the Langevin kicks to each ion's energy so that the overall growth rate of the ion beam's longitudinal emittance is given by the value as calculated from Piwinski's formula[12]. For an ion at its nth circulation with longitudinal coordinate $(\phi_n, P_n)$, after receiving a momentum kick $\delta_P$ due to IBS, its action as defined in eq. (5) increases by

$$\Delta I = I_{n+1} - I_n = \frac{1}{2}\left[(P_n + \delta_P)^2 - P_n^2\right] = P_n \delta_P + \frac{1}{2}\delta_P^2. \quad (B1)$$

Assuming $\delta_P$ is uncorrelated with $P_n$, the average increase of the ion's action is given by the ensemble average of eq. (B1)

$$\langle \Delta I \rangle = \frac{\langle \delta_P^2(\phi) \rangle}{2} . \quad (B2)$$

Assuming

$$\langle \delta_P^2(\phi) \rangle = \delta_{max}^2 \cdot \exp\left(-\frac{\phi^2}{2\sigma_\phi^2}\right) , \quad (B3)$$

the average increase of ion beam's action is

$$\langle \Delta I \rangle = \frac{\langle \delta_P^2 \rangle}{2} = \frac{1}{4\pi I_{ion}} \int_{-\infty}^{\infty} \exp\left(-\frac{I}{I_{ion}}\right) \delta_{max}^2 \cdot \exp\left(-\frac{\phi^2}{2\sigma_\phi^2}\right) dP d\phi = \frac{\delta_{max}^2}{2\sqrt{2}} . \quad (B4)$$

The average action of the beam before the IBS kick is

$$\langle I \rangle = \frac{1}{4\pi I_{ion}} \int_{-\infty}^{\infty} \exp\left(-\frac{I}{I_{ion}}\right)(P^2 + \phi^2) dP d\phi = I_{ion} = \sigma_P^2. \tag{B5}$$

The IBS rise time for the beam energy spread is

$$\frac{1}{T_{IBS}} = \frac{1}{T_{rev}} \frac{\langle \Delta I \rangle}{2 \langle I \rangle}. \tag{B6}$$

Inserting eqs. (B4) and (B5) into eq. (B6) yields

$$\delta_{max}^2 = \frac{4\sqrt{2}\sigma_P^2}{T_{IBS}} T_{rev}. \tag{B7}$$

With the line density of the beam taken as

$$\rho(\phi) = \frac{1}{\sqrt{2\pi}\sigma_\phi} \exp\left(-\frac{\phi^2}{2\sigma_\phi^2}\right), \tag{B8}$$

the standard deviation of the kick should be

$$\langle \delta_P^2(\phi) \rangle = \delta_{max}^2 \cdot \exp\left(-\frac{\phi^2}{2\sigma_\phi^2}\right) = \delta_{max}^2 \frac{\rho(\phi)}{\rho_{max}} = 8\sqrt{\pi}\sigma_\phi \sigma_P^2 \frac{T_{rev}}{T_{IBS}} \rho(\phi). \tag{B9}$$

If we take a random number $X_i$ uniformly distributed around -1 to 1, the amplitude in front the random number should be

$$d_{IBS} = \left[3\langle \delta_P^2(\phi) \rangle\right]^{1/2} = \left[24\sqrt{\pi}\sigma_\phi \sigma_P^2 \frac{T_{rev}}{T_{IBS}} \rho(\phi)\right]^{1/2}. \tag{B10}$$